# First imaging results of a bubble-assisted Liquid Hole Multiplier with SiPM readout in liquid xenon

E. Erdal[a*], A. Tesi[a], D. Vartsky[a], S. Bressler[a], L. Arazi[b] and A. Breskin[a]

[a] *Department of Particle Physics and Astrophysics, Weizmann Institute of Science, Rehovot 7610001, Israel*
[b] *Nuclear Engineering Unit, Faculty of Engineering Sciences, Ben-Gurion University of the Negev, Beer-Sheva 8410501, Israel*

*Email*: eran.erdal@weizmann.ac.il

**Abstract**

First imaging results in liquid xenon of a Liquid Hole Multiplier (LHM) coupled to a Quad-Silicon Photomultiplier (SiPM) array are presented. Ionization electrons deposited in the noble liquid by 5.5 MeV alpha particles, are collected into the holes of a Thick Gas Electron Multiplier (THGEM) electrode having a xenon gas bubble trapped underneath. They drift through the liquid-gas interface, inducing electroluminescence within the bubble. The resulting photons are detected with a Hamamatsu VUV4 quad-SiPM array - providing the deposited energy with a charge-only RMS resolution of 6.6%. The image reconstruction resolution was estimated to be ~200 µm (RMS).

* Corresponding author

## 1. Introduction

Noble-liquid time projection chambers (TPCs) play a central role in neutrino physics experiments and in dark-matter searches [1, 2]. Present-day detectors rely on *single-phase* (liquid) and *dual-phase* (liquid-vapour) concepts. In the former, radiation-induced charges and scintillation photons in the liquid are recorded respectively with charge-sensing electrodes and photon detectors. In the latter, in addition to scintillation recording in the liquid, electrons extracted into the vapour phase are detected either through electroluminescence (EL) [3] or after moderate avalanche multiplication [4].

A recent concept has been proposed and investigated for the combined detection of radiation-induced ionization electrons (S2) and primary-scintillation (S1) photons in noble liquids: the so-called *bubble-assisted Liquid Hole-Multiplier* (LHM) [5- 10]. The LHM consists (Fig. 1) of a perforated electrode, e.g., that of a GEM [11] or a THGEM [12], immersed in the liquid - with a noble-gas bubble trapped underneath; its top surface is optionally coated with a UV- photocathode, e.g. CsI. A plane of heating wires below the electrode generates the bubble that, once formed, remains stable as long as the system is in a thermodynamic steady state [8]. Ionization electrons drifting from the liquid volume towards the electrode are focused into the holes and induce the S2 EL light flash after crossing the liquid-vapour interface. Primary scintillation photons, which conventionally constitute the S1 signal detected directly by photon detectors, are detected in the LHM by inducing the emission of photoelectrons from the CsI surface; these are collected into the holes and cross the liquid-vapour interface into the bubble (at the holes' bottom), where they induce an EL signal defined here S1' (Fig. 1). While in our previous works, the EL photons (up to ~400/e/4π) were detected with a PMT [10], in the present study they are detected and localized with a Quad-SiPM array located below the bubble.

The main potential advantage of the LHM concept is that it may enable maintaining a high resolution in measuring the ionization charge (S2 signal) over large-area noble liquid TPCs. In the standard dual-phase TPC scheme, non-uniformities in S2 arise due to variations in the width of the gap between the gate and anode meshes, as well as because of instabilities of the liquid-vapour interface; these result in considerable fluctuations in the EL yield, which are expected to increase with the size of the TPC. On the other hand, in a large-area noble liquid TPC comprising a tiled array of LHM modules, each with its own bubble, the well-defined location of the liquid-vapour interface at the bottom of the holes, should result in reduced EL fluctuations. Note that the S2 energy resolution obtained so far in our experiments ($\sigma/E = 5.5\%$ for ~7,000 ionization electrons entering the holes [10]) with 3 cm-diameter LHM electrodes; it is roughly twice better compared to that obtained for a similar number of ionization electrons in the 30 cm-diameter XENON100 dual-phase TPC [3]. Evidently, it has to be proven that good S2 resolution can be maintained on large-size LHM modules.

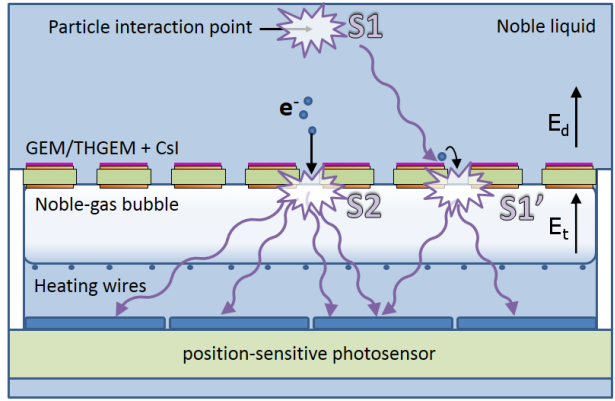

**Figure 1**: Conceptual scheme of the LHM. A perforated electrode (GEM or THGEM) is coated on top with a CsI photocathode; a heating-wires grid forms a bubble underneath; a position-sensitive photon detector (e.g. a SiPM array) is located below. The stable bubble is supported by buoyancy against the bottom of the electrode. Ionization electrons focused into the holes create EL light (S2) once they cross the liquid-gas interface into the bubble. Primary scintillation (S1) photons impinging on the photocathode release photoelectrons which are focused into the holes and create similar EL signals (S1'). The lateral coordinates of the EL signals are reconstructed by the position-sensitive photon detector.

The evaluation of SiPMs in LXe has been performed by several investigators, e.g. [13, 14]. Others have investigated their operation in LAr [15, 16]. Event imaging in a high-pressure Xe TPC using SiPMs is employed by the NEXT collaboration searching for neutrinoless double beta decay in $^{136}$Xe [17]. SiPM arrays can provide much higher granularity (and therefore image reconstruction accuracy) than PMTs, which is the main reason for their use in the present work; this is in fact also necessary for maintaining a high S2 resolution in large LHM modules, to correct for position-dependent effects close to the LHM perimeter. Other advantages making SiPMs attractive for future rare-event experiments in comparison to PMTs include higher radio-purity, inherent compactness (which makes it easier to implement them in a TPC design), lower operational voltage and lower price, while maintaining high gain and single-photon sensitivity.

In this work we present our first results of event localization in a bubble-assisted LHM detector equipped with a Quad-SiPM array immersed in LXe.

## 2. Experimental setup & methodology

The present experiments were carried out in our MiniX LXe cryostat; its structure and operation were described in detail in [8]. The LHM detector (Fig. 2) had a 0.4 mm thick bare THGEM electrode with a hexagonal holes' pattern (0.3 mm in diameter holes, spaced by 0.7 mm), without a photocathode; therefore, the detector was sensitive only to ionization electrons. The latter were deposited by alpha particles from a homemade $^{241}$Am source (activity of ~190 Bq); it had a ~0.5 mm broad annular shape, of ~3.9 mm in diameter (see source image in the results section below). A grid of parallel heating wires (Ni-Fe, 55 µm in diameter, 2 mm spacing) was placed 1.5 mm below the electrode, generating the bubble. Polarized at an adequate potential, the wire-grid established an electric field across the bubble (transfer field, $E_t$); its intensity permitted the generation of EL signals close to the liquid-gas interface at the hole's bottom, across the bubble (in the transfer gap) and at the vicinity of the wires.

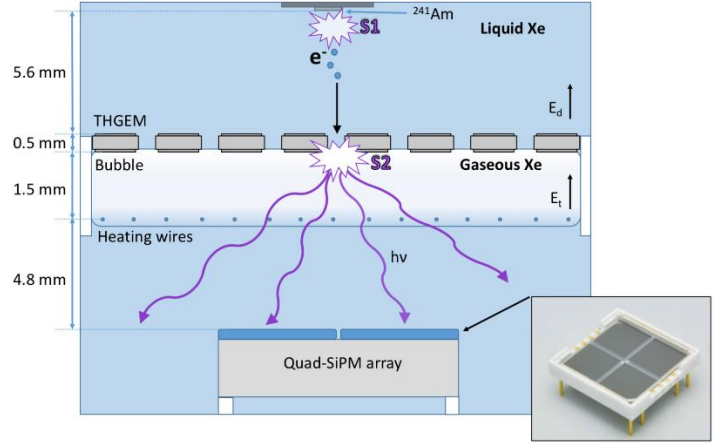

**Figure 2**: The Quad-SiPM LHM detector scheme (not to scale). A 0.4 mm thick THGEM electrode (with 0.3 mm in diameter holes spaced by 0.7 mm) is immersed in LXe; the gas bubble underneath is formed by a grid of heating wires, spaced 2 mm apart, located 1.5 mm below the electrode; the Quad-SiPM array is located at 4.8 mm under the wire grid. Ionization electrons focused into the holes induce EL light (S2) in the bubble; a fraction of the primary scintillation S1 flash traverses the holes. Both photon signals are detected by the SiPMs, yielding the event's energy signal and location. The 12 x 12 mm$^2$ Quad-SiPM detector is shown on the right.

The EL signals were recorded with a quad-SiPM array located in the liquid, 6.3 mm under the THGEM electrode's bottom face. This Hamamatsu Quad VUV4 MPPC (model S13371-6186) is suitable for operation at cryogenic temperatures; with its quartz window, it is sensitive to the Xe VUV excimer photons (175 nm). Each SiPM segment has an area of 6x6 mm$^2$, with a 0.5 mm gap between segments; it has 13,923 pixels per segment and a geometrical fill-factor of ~60%. It was mounted on a printed board, with the R-C supply circuitry (shown in Fig. 3). The operation voltage was maintained at -57 V; the photon detection efficiency (PDE) is ~15% at 175 nm, as stated in [18].

Electroluminescence-induced signals from each segment were digitized with a Tektronix digital oscilloscope (MSO 5204B) and recorded for offline post-processing using dedicated Matlab scripts. The trigger was provided by primary-scintillation (S1) photons reflected off PTFE spacers (not shown; located above around the source holder, for further details see [10]) and detected by a PMT located above the $^{241}$Am source (not shown). The choice of S1 as a trigger is natural in this specific setup for two technical reasons (a) primary scintillation from 5.5 MeV alpha particles generates a large prompt flash of light (~10$^5$

photons/4π) and (b) the fixed drift distance of electrons from the interaction point to the detector facilitates analyzing the time structure of their induced pulse shape.

Ionization electrons deposited by the 5.5 MeV alpha articles (40 µm range in LXe), were drifted under $E_d$ towards the THGEM-electrode and focused into its holes. The EL induced at the bottom of each hole permitted imaging the holes' pattern and to determine the position resolution of the reconstruction technique. Event position reconstruction was done by a simple center-of-gravity (COG) method, followed by calibrating the distances according to the electrode's holes pattern. Mathematically:

$$\vec{R}_i = A \cdot \frac{\sum_j \vec{r}_j L_{ij}}{\sum_j L_{ij}}$$

Where $\vec{R}_i$ is the reconstructed location of event $i$, $L_{ij}$ is the light intensity (i.e. the integral under the recorded pulse) on pad $j$ at event-number $i$, $\vec{r}_j$ is the location of the center of the $j$'s pad and $A$ is a global scaling factor. The latter was used for maintaining the 0.7 mm spacing between the hole centers. Finally, a 2D histogram of the reconstructed locations was plotted.

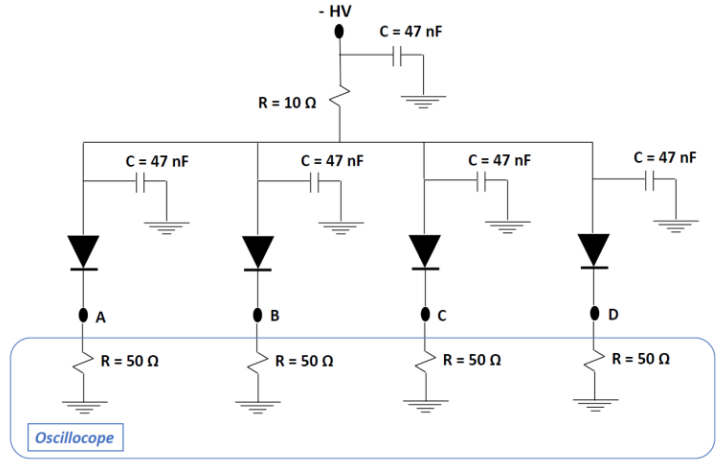

**Figure 3**: Electrical diagram of the Quad SiPM supply board and readout (50 Ω at the oscilloscope input).

### 3. Results

*3.1 Analysis of SiPM signals*

A typical few-photons spectrum, recorded in LXe (T~173 K) on a single SiPM segment is shown in Fig. 4. The waveforms were recorded by setting a low trigger on the SiPM output; the signals are due to dark pulses originating from the SiPM with some contribution by pixel-cross talk.

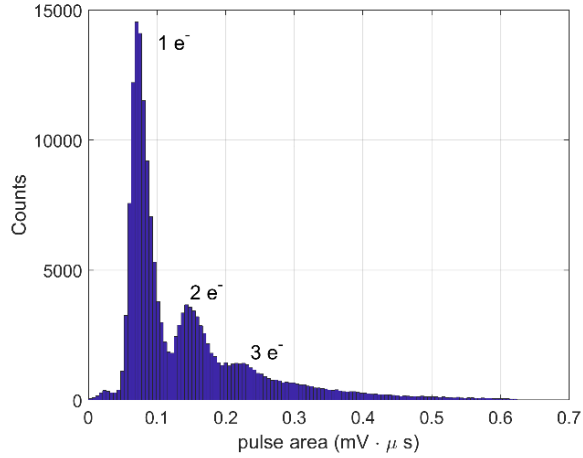

**Figure 4**: A dark-noise spectrum recorded in LXe, on a single SiPM segment, showing the response to one or a few electrons.

An example of typical alpha-particle waveforms recorded (in the detector shown in figure 2) from each of the four Quad-SiPM channels is shown in figure 5. One can observe a small fraction of the primary-scintillation S1 photons that traversed the THGEM holes, and the S2 EL signals. The waveforms shown in figure 5a were recorded at a voltage applied across the THGEM electrode $\Delta V_{THGEM} = 2.1$ kV, with $E_t = 0$ and $E_d = 0.5$ kV/cm; they correspond to EL produced close to the hole's bottom; the total pulse width at the base is ~0.5 microseconds. The ones shown in figure 5b were recorded at the same $\Delta V_{THGEM}$ and $E_d$ values but with $E_t = 12$ kV/cm; their shape is due to EL photons generated at the hole's bottom (first peak), along the electrons' path across the bubble (middle dip) and in the high-field region approaching the grid-wires (second peak); the total pulse width at the base ~2 µs.

The S2 energy resolution of the present SiPM-LHM setup was derived from the distribution of the number of photoelectrons recorded by the SiPM - computing the area under each waveform, normalized to that of a single photoelectron. An example of a pulse-area spectrum is shown figure 6, for LHM voltage settings: $\Delta V_{THGEM} = 3.6$ kV, $E_d = 0.5$ kV/cm and $E_t = 0$; this distribution, induced by alpha particles, was recorded from the four SiPM segments. A Gaussian fit provides the distribution mean (µ) and resolution. Similar to previous S2 resolution (~7% RMS) obtained with a THGEM-LHM and a PMT readout, the energy resolution obtained here with the SiPMs is 6.6% RMS. Note that the addition of a transfer field resulted in a slight degradation of the energy resolution (to 9% RMS); this however may be improved by optimizing the detector operation parameters, as discussed below. The low-energy tail of the distribution in figure 6 is attributed to alpha particles leaving a fraction of their energy in the source's matrix, while the excess of events at the right of the peak corresponds to the coincidence of 5.48 MeV alphas with the source's 60 keV gammas (as discussed in [10]).

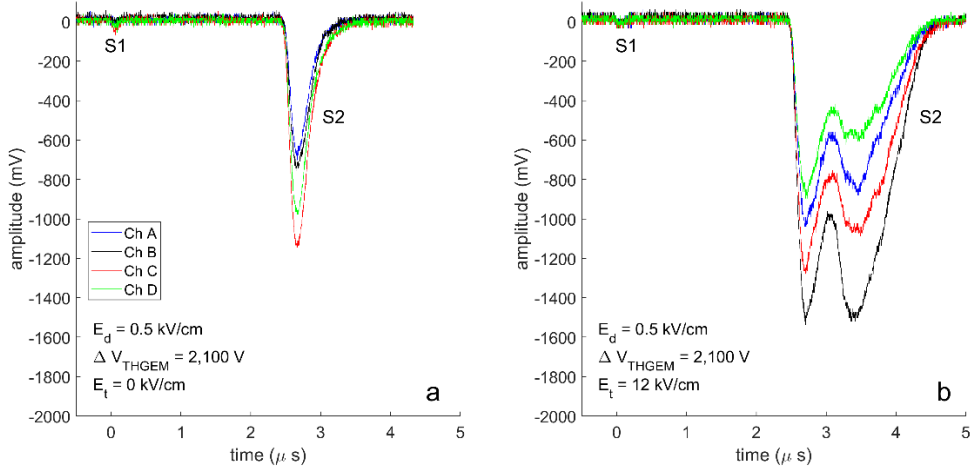

**Figure 5**: a) Typical waveforms recorded on the four SiPM pads (A-D) with no transfer field ($E_t = 0$) showing residual S1 pulses and S2 EL pulses generated at the vicinity of the liquid-bubble interface at the bottom of the hole. b) Waveforms recorded under a relatively high transfer field ($E_t = 12 \text{ kV/cm}$), showing S1 pulses and S2 EL pulses from the vicinity of the liquid-bubble interface, along the transfer gap and close to the heating wires.

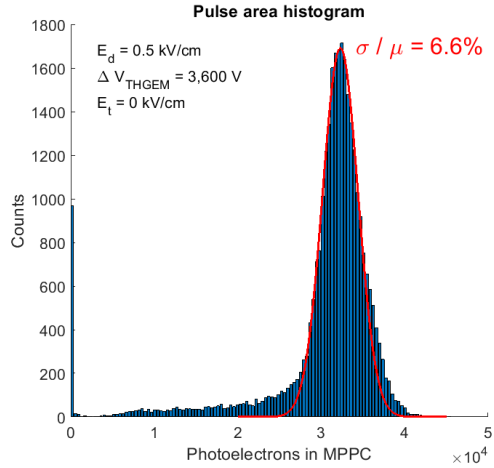

**Figure 6**: Energy spectrum (pulse-area distribution) obtained by summing the Alpha particles' EL-induced charge, recorded on all four SiPM pads.

## 3.2 Imaging with the Quad-SiPM array

In the detector geometry shown in figure 2, electrons are deposited within a few tens of µm from the surface of the alpha source; each event-induced electron-swarm drifts along the field lines (under $E_d$) towards the nearest hole (and splits rarely into two or three holes). The expansion of the electron swarm due to diffusion during its drift from the interaction point to the LHM electrode is negligible (in this particular setup) as the RMS deviation is 80 µm [19]. Longitudinal diffusion is one order of magnitude smaller. Therefore we expect obtaining an image of the hole pattern with a general shape of the ring-shaped alpha source (shown in figure 7a). The EL occurs along the entire electrons' drift path in the bubble. At zero transfer field, the electrons' trajectories, after crossing the liquid-gas interface at the bottom of hole, are deviated along the field lines towards the electrode's bottom face. This lateral movement causes EL-photons emission along their entire path, thus somewhat blurring the holes' image. Therefore, to obtain a well-resolved image, we have proceeded as follows: a) an intense transfer field was applied, causing the electrons to drift within the bubble towards the wires grid, and b) performing the COG computation only for the first part of the EL emission, that originating from the vicinity of the liquid-bubble interface. The hole pattern shown in figure 7b, is a result of computing the COG over the first 450 ns (hole-bottom vicinity) of the pulses shown in figure 5b. It reproduces well, qualitatively, the shape and some details of auto-radiographic source image (figure 7a) recorded by a Fuji phosphor-imager scanner (model FLA-9000, plate model BAS-TR2040S). The 4.5 x $10^5$ waveforms forming our SiPM-LHM image were recorded (over ~5 hours) with $E_d = 0.5$ kV/cm, $\Delta V_{THGEM} = 2.1$ kV and $E_t = 12$ kV/cm. The THGEM-electrode holes are clearly apparent, with their reconstructed locations of 0.7 mm spacing between their centers.

As explained above, the overall image granularity is dictated by the holes spacing and their diameter, however the resolution of the COG position reconstruction can be determined by observing the light distribution within an individual hole. We consider for the purpose of this analysis the light of a single hole as a point source, which will give an upper bound on the distribution width. More advanced analysis methods are the subject of further systematic studies. The projected COG distribution across the x-axis of the encircled hole in figure 7b is depicted in figure 7c. This hole was chosen, being well separated from the neighboring ones (probably due to the interplay between the hole pattern and the activity pattern of the source). The COG distribution was fitted with a Gaussian, yielding a resolution of ~200 µm RMS.

**Discussion**

In this article we have shown, for the first time, the imaging capability of the bubble-assisted LHM detector with a pixelated SiPM-array readout in LXe. A ~3.9 mm diameter ring-shaped [241]Am source was imaged with a LHM comprising a THGEM-electrode with 0.3 mm in diameter holes, viewed by a 2×2 array of 6×6 mm² SiPM elements. The resulting image, reconstructed from the COG of EL photons emitted close to the liquid-gas interface, reproduced the ring shape of the source and the holes pattern (0.7 mm center-to-center) of the electrode. The position reconstruction resolution of EL-photons emitted from a single hole is ~200 µm RMS. The pulse-area S2 energy resolution derived from all holes is 6.6% RMS. It is of the same order as that measured in previous works with a PMT, under similar conditions.

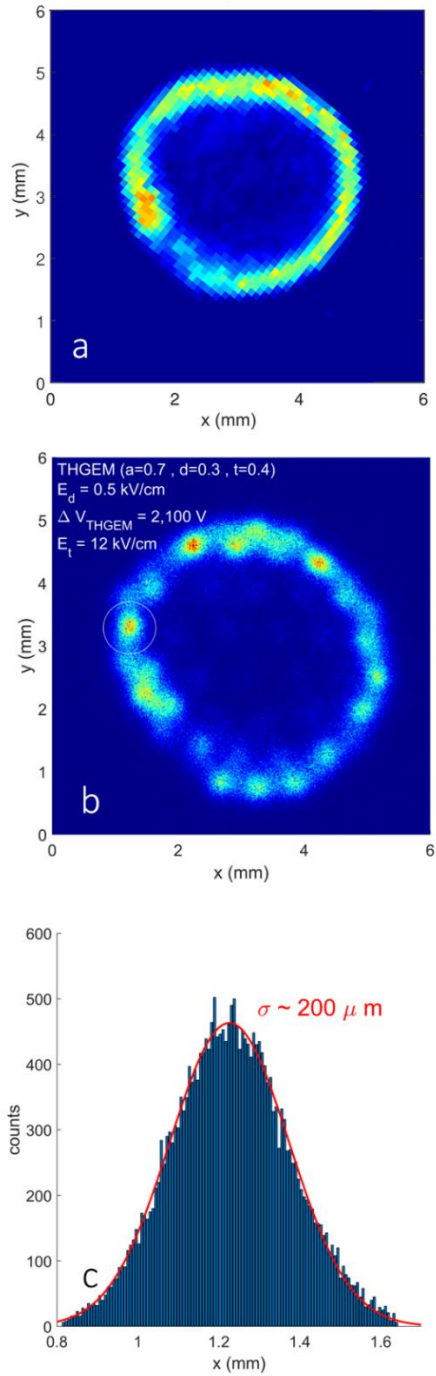

**Figure 7**: (a) the auto-radiographic Alpha-source image, recorded with a Fuji phosphor-imaging plate; (b) 2D histogram of the EL photons emitted at the vicinity of the liquid-gas interface, recorded with the Quad-SiPM LHM detector; the holes' pattern of the THGEM electrode is well resolved; (c) projected COG distribution across the x-axis, of the encircled hole in b), with a Gaussian fit to the data.

These first encouraging results call for further investigations of the electrodes geometry and that of SiPM elements, operation parameters, and COG-derivation algorithms.

As stated above, the overall image granularity depends on the hole pattern, which acted here as a "lens" for the source-induced ionization electrons in LXe. Thus, the use of electrodes with finer hole diameter and pitch (e.g. GEM-LHM [10]) would result in smaller deviation of the event-deposited electrons from their original location; this will naturally enhance the event's location reconstruction.

The proof-of-principle of SiPM-LHM imaging capability was performed with alpha particles and a specific set of electric-field values, chosen quite arbitrarily. As an example, although demonstrating good position resolution in certain conditions, employing only part of the EL-photons flash, the best energy-resolution has been achieved at higher potential difference across the THGEM electrode; it thus calls for better optimization of the parameters. Specifically, the drift field, the voltage across the electrode and transfer field all play a crucial role in determining these parameters and require careful studies. One should also replace the present wire grid with a finer spaced one, a fine mesh or even a conductive window – to avoid electron-path distortions while drifting within the bubble.

The conventional COG method employed here, though being the simplest one, is far from being ideal for event-location reconstruction. It was quasi-linear at the center of the present small-area Quad-SiPM array, up to half the size of an individual segment, distorting the image at larger distances. Other methods such as iterative position weighed COG algorithms [20, 22] or statistical methods, such as Maximum Likelihood or Least Squares algorithms [22, 23] are expected to provide better linearity and a more robust position reconstruction. These will be part of future studies, with larger SiPM arrays and optimized inter-element spacing.

The study performed here was done with 5.5 MeV alpha particles, yielding in this particular setup ~100 photons/e/$2\pi$ emitted from the bubble. In the next steps, we plan investigating the achievable localization properties with lower-ionization sources, e.g. low-energy x-rays; these may require higher light amplification in the LHM element. Such studies are of high relevance for evaluating the SiPM-LHM applicability in future large-volume rare-event experiments.

**Acknowledgements**

This work was partly supported by the Israel Science Foundation (Grant No. 791/15). The research was carried out within the R&D program of the DARWIN Consortium for future LXe dark matter observatory. It is part of the CERN-RD51 detector R&D program.